\documentclass[11pt,twoside]{article}
\usepackage{./asp2014}

\aspSuppressVolSlug
\resetcounters

\begin{document}

\title{Measurement of the evolution of the magnetic field of the quiet
  photosphere during a solar cycle}
\author{Renzo Ramelli,$^1$ Michele Bianda,$^1$ Svetlana Berdyugina,$^2$ Luca
  Belluzzi,$^{1,2}$ and Lucia Kleint$^3$
\affil{$^1$Istituto Ricerche Solari Locarno IRSOL, associated to Universit\`a della Svizzera
  Italiana, Locarno, TI, Switzerland;
}
\affil{$^2$Kiepenheuer-Institut f\"ur Sonnenphysik, Freiburg, Germany; 
}
\affil{$^3$University of Applied Sciences Northwestern Switzerland FHNW,
  Windisch, AG, Switzerland 
}}

\paperauthor{Renzo Ramelli}{ramelli@irsol.ch}{0000-0002-1976-1024}{Universit\`a della Svizzera Italiana}{Istituto Ricerche Solari Locarno, IRSOL}{Locarno}{TI}{6605}{Switzerland}
\paperauthor{Michele Bianda}{mbianda@irsol.ch}{}{Universit\`a della Svizzera Italiana}{Istituto Ricerche Solari Locarno, IRSOL}{Locarno}{TI}{6605}{Switzerland}
\paperauthor{Svetlana Berdyugina}{sveta@leibniz-kis.de}{0000-0002-2238-7416}{Kiepenheuer-Institut f\"ur Sonnenphysik}{}{Freiburg}{}{79104}{Germany}
\paperauthor{Luca Belluzzi}{belluzzi@irsol.ch}{}{Universit\`a della Svizzera Italiana}{Istituto Ricerche Solari Locarno, IRSOL}{Locarno}{TI}{6605}{Switzerland}
\paperauthor{Lucia Kleint}{lucia.kleint@fhnw.ch}{0000-0002-7791-3241}{University of Applied Sciences Northwestern Switzerland FHNW}{Institute of 4D Technologies}{Windisch}{AG}{5210}{Switzerland}

\begin{abstract}
The solar photosphere is filled by a magnetic field which is tangled on scales
much smaller than the resolution capability of solar telescopes. 
This hidden magnetic field can be investigated via the Hanle effect. In 2007
we started a synoptic program to explore if the magnetic flux of the quiet
photosphere varies with the solar cycle. For this purpose we applied a differential Hanle
effect technique based on observations of scattering polarization in C$_2$
molecular lines around 514.0 nm, taken generally every month. Our results now
span almost one complete solar cycle.
\end{abstract}

\section{Introduction}

The solar photosphere is seething with a vast amount of magnetic flux tangled
on scales much smaller than the resolution that can be achieved by 
solar telescopes. Observing techniques based on the Zeeman effect are
in practice blind to such magnetic flux due to signal cancellation effects
generated by mixed polarity fields present in a single resolution element. 
This hidden magnetism can be however investigated considering the Hanle effect
observed with high precision spectropolarimetric measurements. 

In 2007, when the minimum
of the solar cycle was approaching, 
we started a synoptic program \citep{Kleint11,Kleint10} to
explore possible variations with the solar cycle of this tangled magnetic
flux. The observations consist in the measurement of 
the scattering polarization of C$_2$ molecular
lines in the spectral region around 514.0 nm in the quiet Sun. 
The magnetic flux is then inferred 
applying 
a differential Hanle effect technique as described by \citet{Stenflo98}.

\section{The observations}

The observations are generally carried out with the cadence of about one month
at the Gregory-Coud\'e telescope in Locarno. With the ZIMPOL polarimeter
\citep{zimpol-spie2010,gandorferetal04} we achieve a polarimetric
resolution of the order of $10^{-5}$, thanks to a fast piezo-elastic modulator (42
kHZ) and a synchronous demodulation, done with a special dedicated
CCD-camera. As a result of the fast modulation, seeing induced effects are
avoided. The scattering polarization measurements are obtained with the
spectrograph slit set parallel to the solar limb at a distance of about 5-10
arcsec from it (i.e, $\mu = \cos\theta =$ 0.1- 0.15) and subtends about 
180 arcsec. A set of five observations is usually carried out in five
heliographic positions: North Pole (N) , North-West at +45$^\circ$ latitude
(NW), West equator (W),
South-West at -45$^\circ$ latitude (SW) and South Pole (S). Including all calibrations
(polarimetric, dark, flat-field), this set of observations takes about half a
day.   

Since beginning of this synoptic program, several improvements were applied to
the instrumentation, such as the deployment of the version III of the
ZIMPOL system in 2010 \citep{zimpol-spie2010} or a new silver coating of the
mirrors. Thanks to them it was possible to increase the observing efficency. A
limb tracking system and other automatic systems allowed to fully automatize
the observing procedure.

\section{Results}

An example of the linear polarization profile (Q/I) obtained in a typical
observation, is shown in Figure \ref{fig1}. The zero level of polarization has
been shifted here in order to correspond to the average  polarization measured in the continuum.
The four marked peaks represent the R1, R2,
R3 and P spectral lines of two different C$_2$ triplets considered in this synoptic
program.

\articlefigure[width=0.8\textwidth]{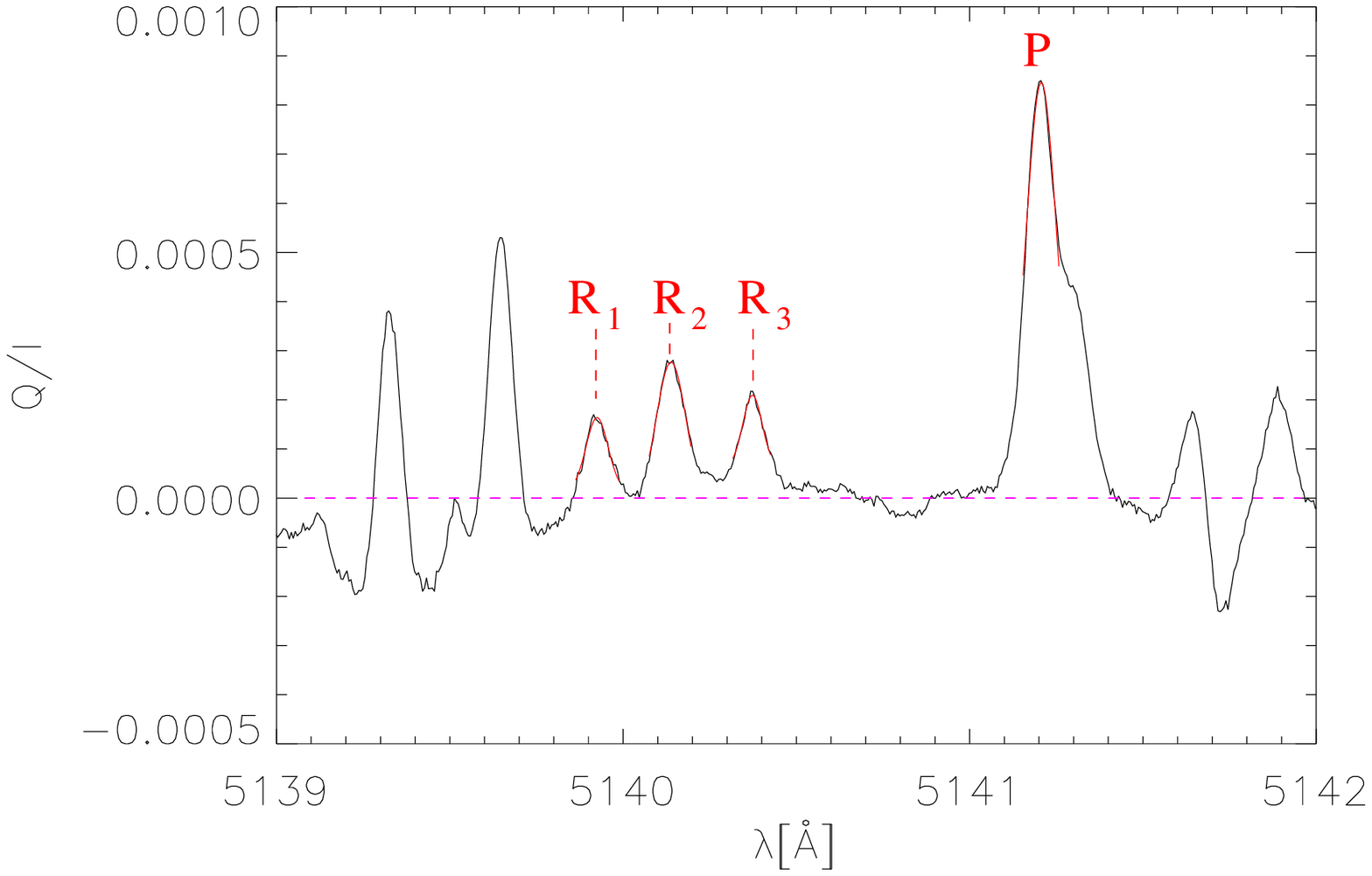}{fig1}{Typical
  linear scattering polarization observation (Stokes Q/I) of C$_2$ spectral lines
  obtained near the solar limb.  (This observation example has been taken on 11th  January 2012 at the West limb).}

\begin{figure}[hbt]
\begin{center}
\includegraphics[width=0.48\linewidth]{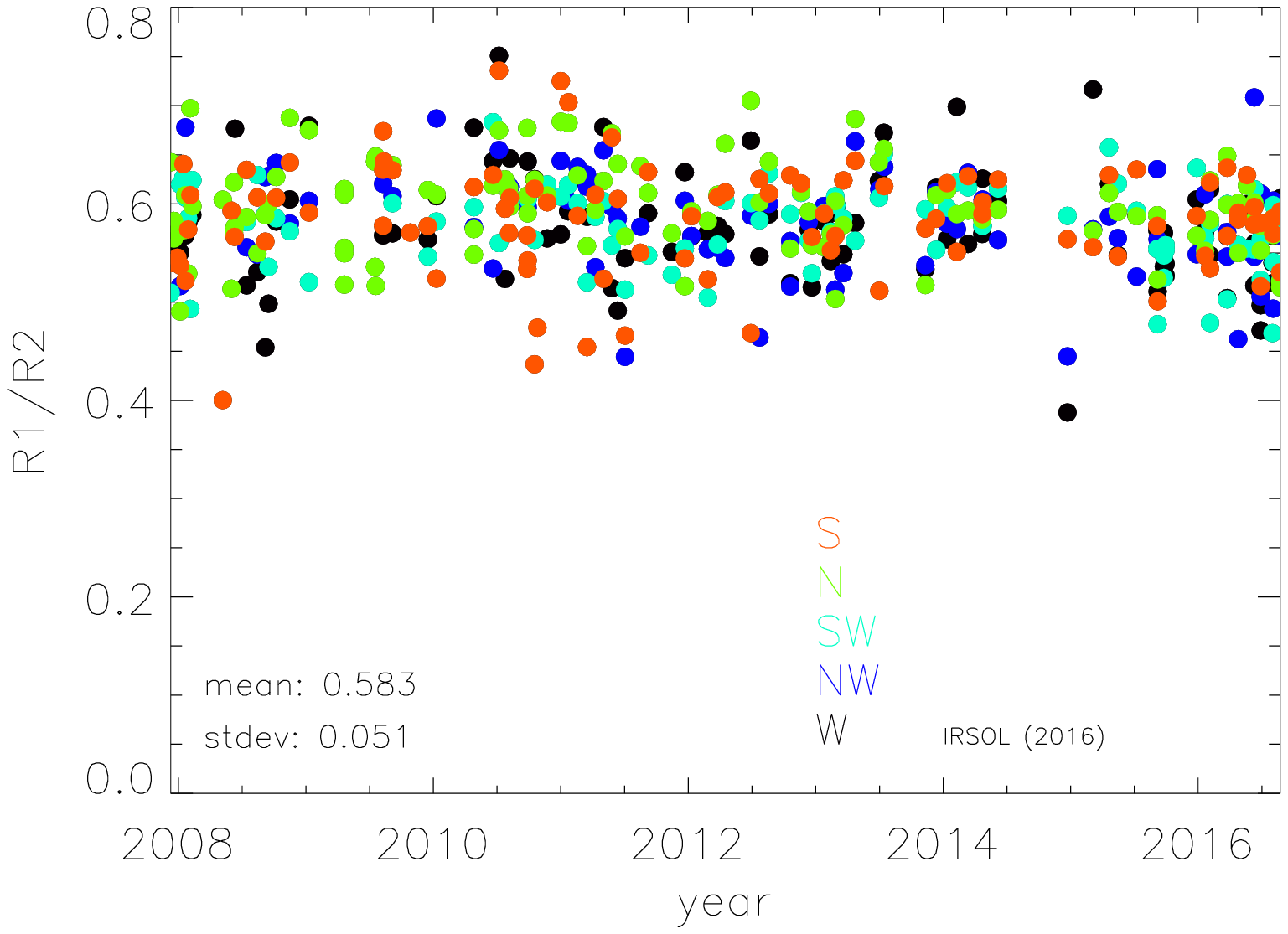}
\includegraphics[width=0.48\linewidth]{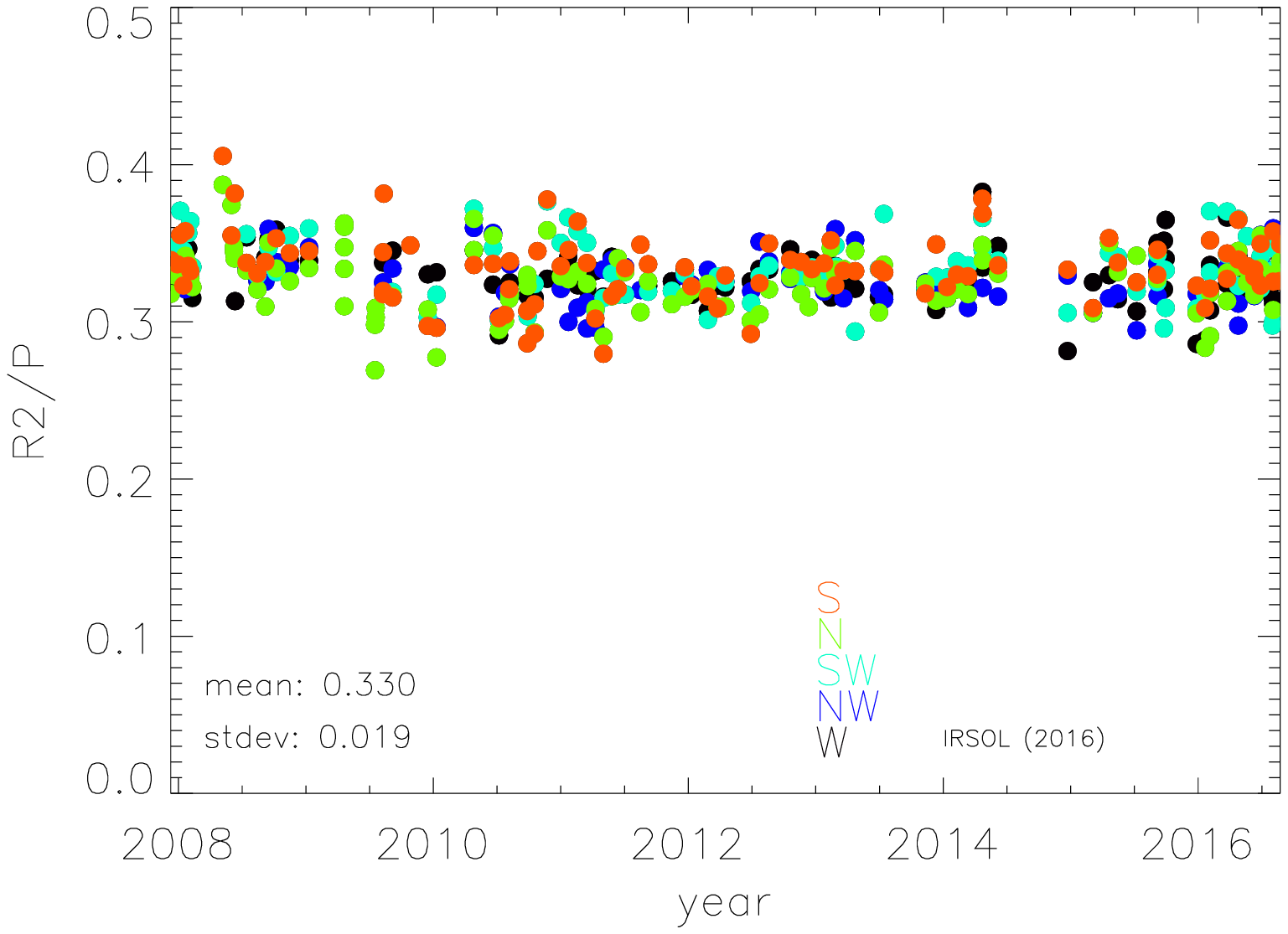}
\includegraphics[width=0.48\linewidth]{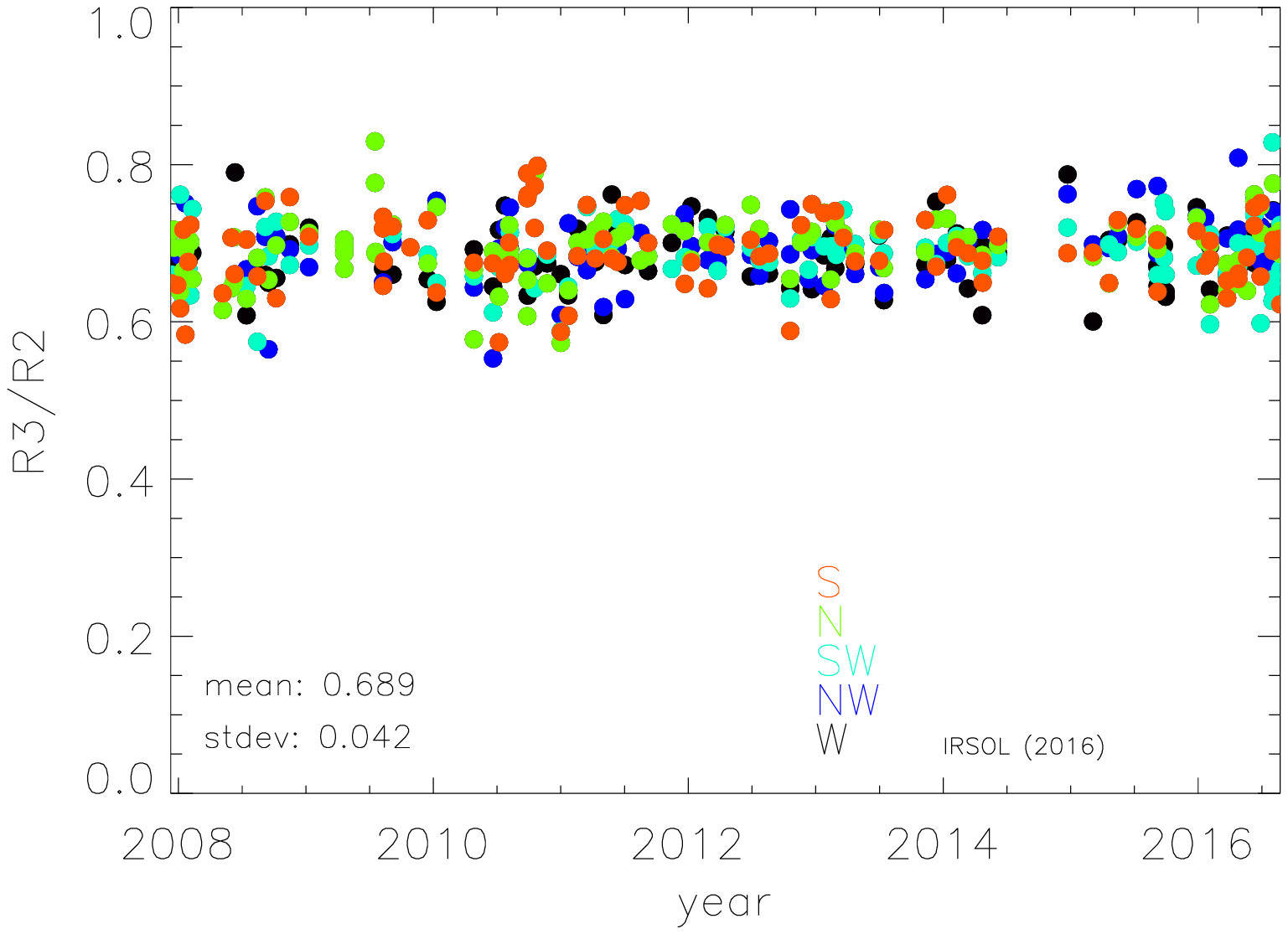}
\caption{\label{fig2}Ratios between the Q/I peaks of the different C$_2$ lines as a
  function of time. The different limb positions are shown with different colors.}
\end{center}
\end{figure}

The temporal evolution of the observed ratios between the Q/I signal
amplitudes of different C$_2$ lines with respect to the continuum 
are reported in Figure \ref{fig2} for the whole
data set.  The
observing period from 2007 to 2016 includes both the minimum (2009) and the
maximum (2014) of the present solar cycle. From the temporal evolution it is
difficult to see significant variations. 

\begin{figure}[hbt]
\begin{center}
\includegraphics[width=0.48\linewidth]{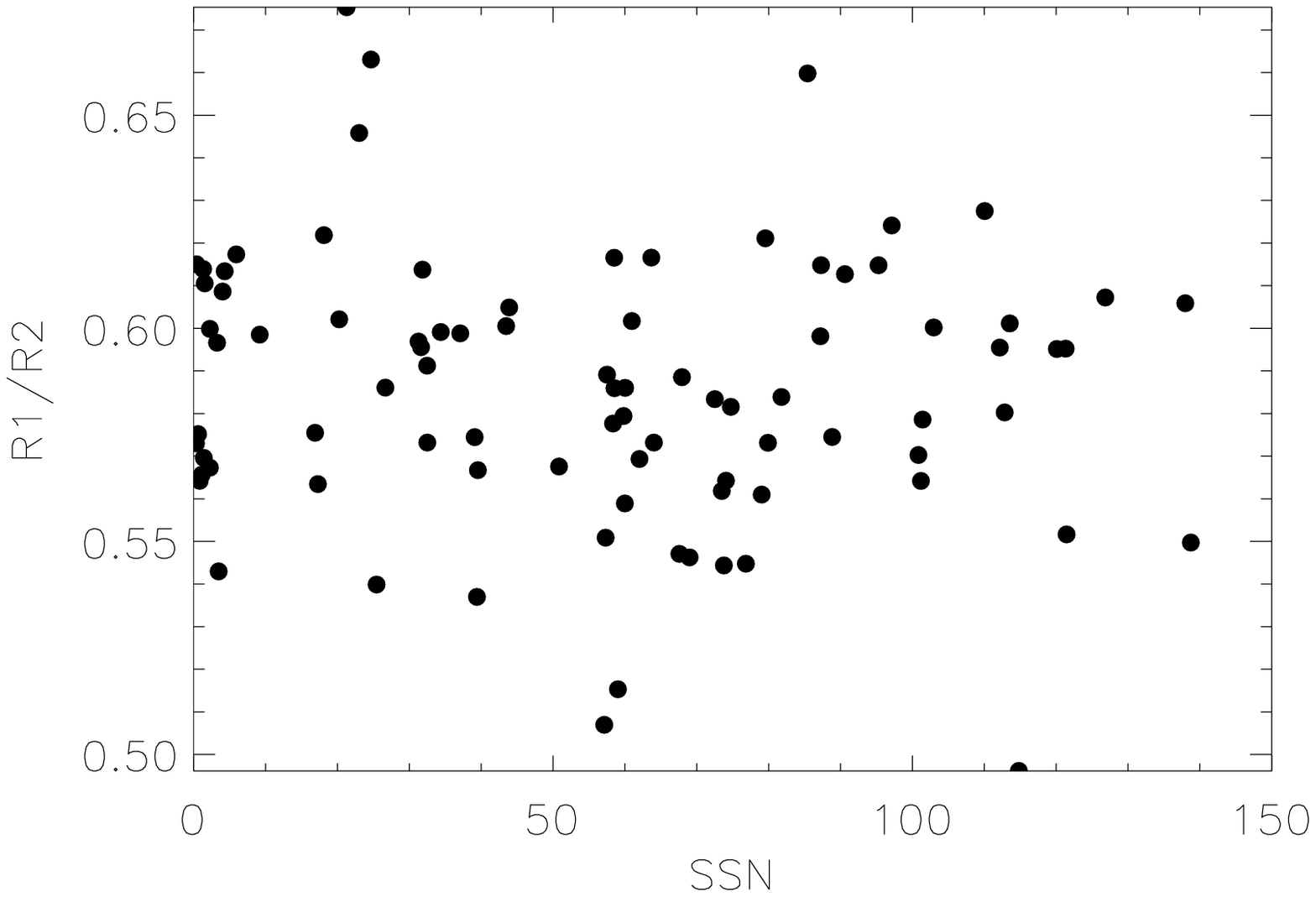}
\includegraphics[width=0.48\linewidth]{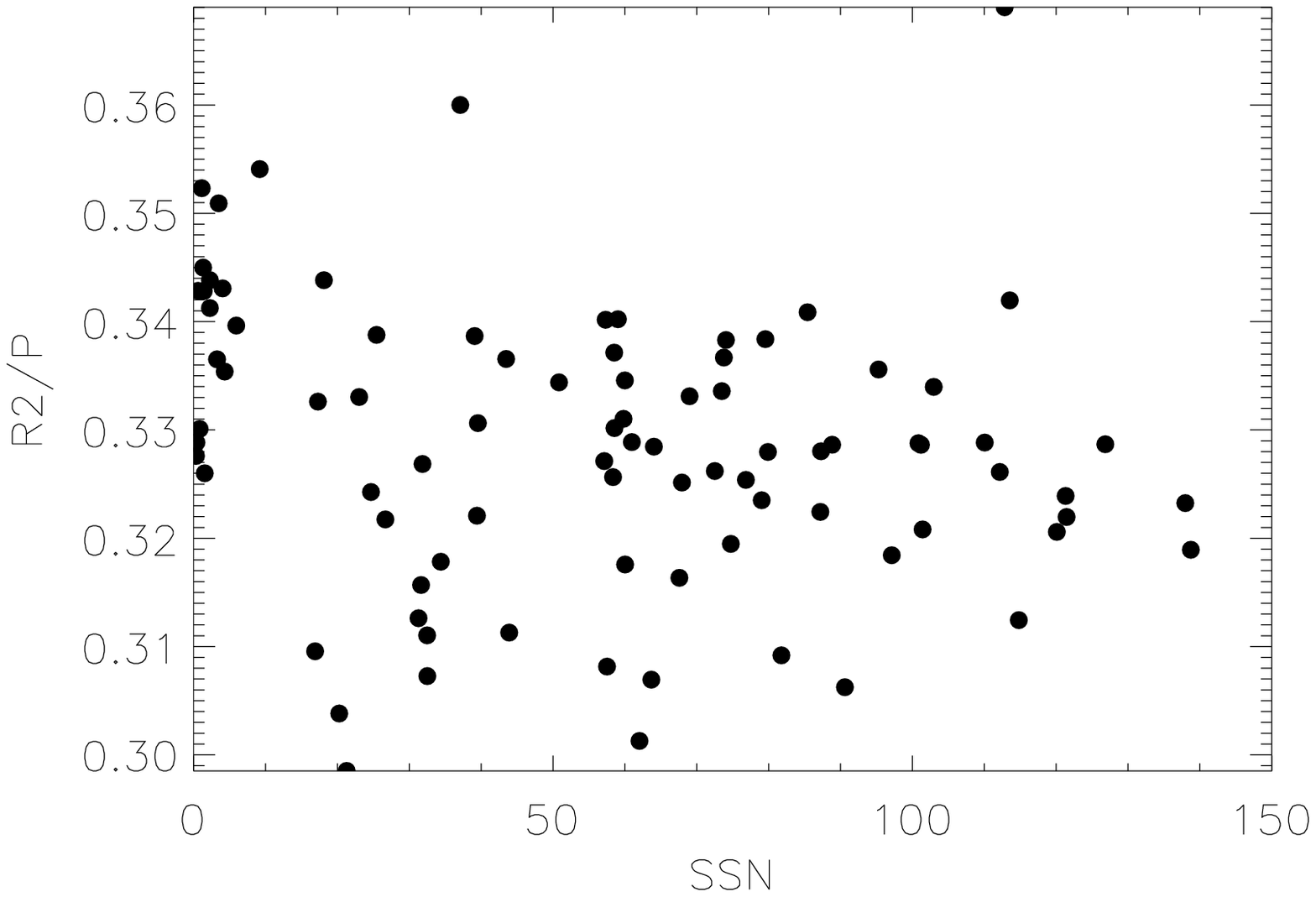}
\includegraphics[width=0.48\linewidth]{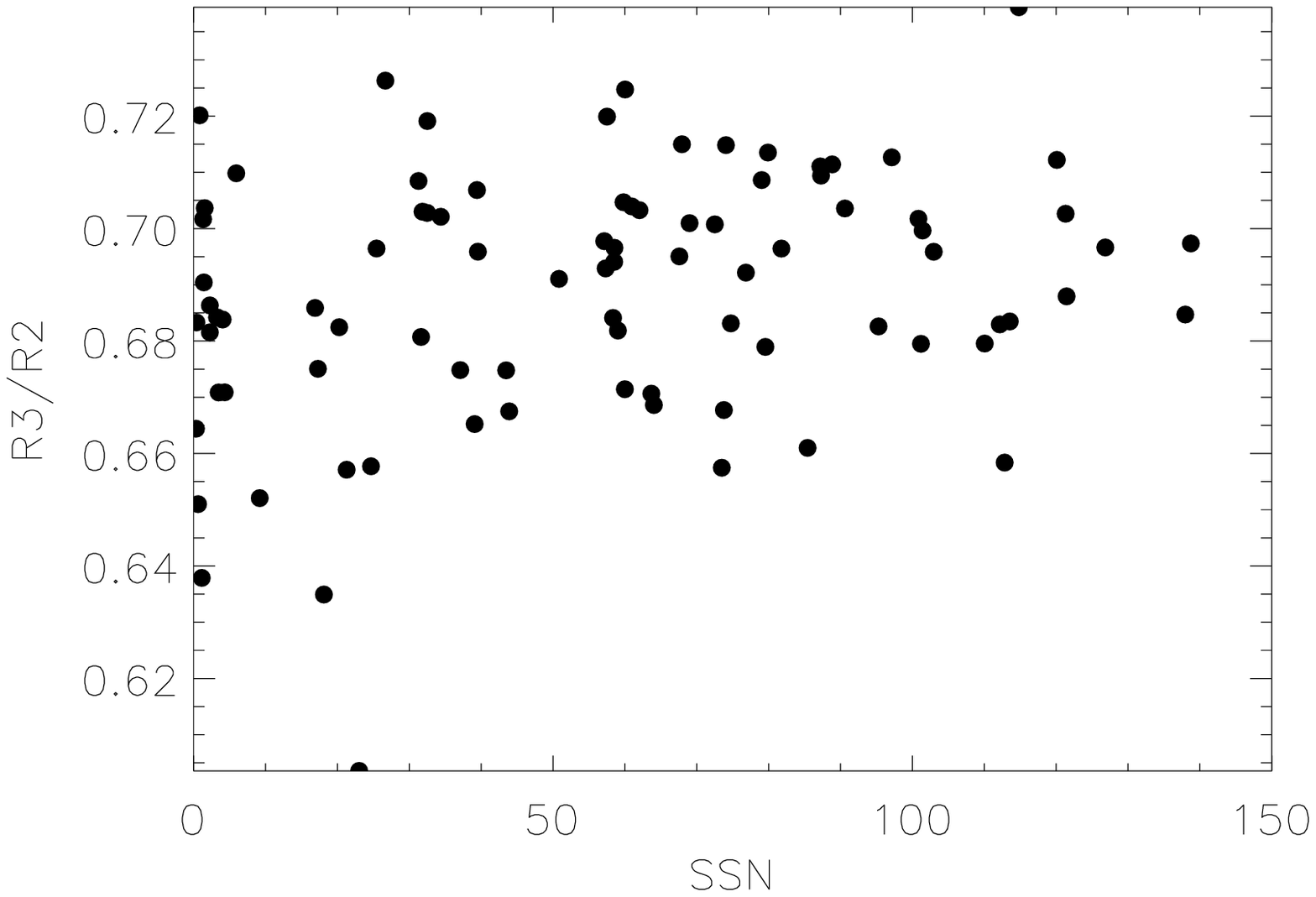}
\caption{\label{fig3}Daily average ratios between the Q/I peaks of the different C$_2$ lines versus the Sunspot Number.}
\end{center}
\end{figure}

In order to better analyze the correlation between the solar cycle and the
observed ratios trough a more careful analysis we report in Figure \ref{fig3}
scatter plots of the measured ratios towards the smoothed International
Sunspot Number (SSN)\footnote{Source: WDC-SILSO, Royal Observatory of Belgium,
Brussels}, which is an index directly related with the amount of solar activity. 
In order to reduce the scattering of the points we average for each
observing day the ratios obtained at the different limb positions. We report
one point on the scatter plots for each day during which we obtained at least
three complete observations at different limb positions. Thus the statistics
include 87 observing days.
The smoothing of the
SSN has been obtained considering the running average over a period of 30 days.


\section{Conclusion}

It is not easy to appreciate long term variations of the measured ratios
among the polarization singals observed in C$_2$ lines. The available 
data do not show a very clear correlation with solar activity, that could lead us to think that the turbulent
unresolved magnetic field strength varies with the solar
cycle. A more detailed and carefull 
interpretation and analysis of the results are in
progress. The observing program will continue and provide more statistics. 
The data will soon cover a complete solar cycle.

\acknowledgements 
IRSOL is supported by the Swiss Confederation (SEFRI) , Canton Ticino, the city of Locarno and the local municipalities.
This research work was financed by SNF grants 200020\_157103  and  
200020\_169418. We are grateful to the Fondazione Aldo e Cele Dacc\`o for their
financial contribution.

\bibliography{ramelli_1}

\end{document}